\documentclass{PoS}
\usepackage[authoryear,round]{natbib}
\title{Radio investigation of Ultra-Luminous X-ray Sources in the SKA Era}

\ShortTitle{ULXs in the SKA Era}

\author{
\speaker{Anna Wolter}$^1$, 
Anthony P. Rushton$^{2,3}$,
Mar Mezcua$^{4}$,
David Cseh$^{5}$,
Fabio Pintore$^{6}$,
Isabella Prandoni$^{7}$,
Zsolt Paragi$^{8}$,
Luca Zampieri$^{9}$
\\ 
$^1$INAF-Osservatorio Astronomico di Brera, Milano; 
$^2$University of Oxford;
$^3$University of Southampton;
$^4$Instituto de Astrofisica de Canarias (IAC);
$^5$Radboud University, Nijmegen;
$^6$Universit\`a degli Studi di Cagliari;
$^7$INAF-IRA Bologna;
$^8$JIVE, Dwingeloo;
$^9$INAF-Osservatorio Astronomico di Padova
\\
E-mail: \email{anna.wolter@brera.inaf.it}
}

\abstract{A puzzling class of exotic objects, which have been known about for more than 30 years, is reaching a new era of understanding. We have discovered hundreds of Ultra Luminous X-ray sources (ULXs) - non-nuclear sources with X-ray luminosity in excess of the Eddington luminosity for "normal" size stellar Black Holes (BH) - and we are making progresses towards understanding their emission mechanisms. 
The current explanations imply either a peculiar state of accretion onto a stellar size BH or the presence of an intermediate mass BH, the long-sought link between stellar and supermassive BHs. Both models might co-exist and therefore studying this class of object will give insight into the realm of accretion in a variety of environments and at the same time find look-alikes of the primordial seed BHs that are thought to be at the origin of todays supermassive BHs at the centre of galaxies.
The radio band has been exploited only scantily due to the relative faint fluxes of the sources, but we know a number of interesting sources exhibiting both extended emission (like bubbles and possibly jets) and cores, as well as observed transient behaviour. The new eras of the SKA will lead us to a major improvement of our insight of the extreme accretion within ULXs. We will both investigate in detail known sources and research new and fainter ones. When we have reached a thorough understanding of radio emission in ULX we could also use the SKA as a discovery instrument for new ULX candidates. The new array will give an enormous space to discovery: sources like the ones currently known will be detected in a snapshot up to 50 Mpc instead of at 5 Mpc with long, pointed observations. 
}

\FullConference{
Advancing Astrophysics with the Square Kilometre Array\\
June 8-13, 2014\\
Giardini Naxos, Sicily, Italy}

\newcommand{\skipthis}[1]{}
\newcommand\nar{New Astronomy Reviews}
\newcommand\apj{ApJ}
\newcommand\apjl{ApJL}
\newcommand\apjs{ApJS}
\newcommand\aap{A\&A}
\newcommand\mnras{MNRAS}
\newcommand\araa{ARA\&A}
\newcommand\nature{Nature}
\newcommand\sci{Science}
\newcommand\an{AN}
\newcommand\pasa{PASA}
\newcommand\rmx{RMxAC}

\begin{document}
\bibliographystyle{plainnat}

\section{Ultra Luminous X-ray Sources}
X-ray observations of nearby galaxies show a population of point-like, off-nuclear sources with luminosities (if isotropic) in excess of the
classical Eddington limit for accretion onto a 10 M$_{\odot}$ compact object, that we categorize as Ultra-Luminous X-ray sources \citep[ULXs; see, e.g.,][for reviews]{Fabbiano06, FS11}. Nowadays, hundreds of ULX candidates have been detected and many of them have been studied in detail \citep[see e.g.][]{RW00, CP02, swartz04, LB05, LM05, walton11, swartz11}.  As the definition of ULX is purely empirical, the class is likely a mixed bag of sources, powered by different mechanisms. We already know that a fraction of candidates in the catalogs  \citep[see e.g.][]{swartz11} are identified with known supernovae or background AGNs \citep[see e.g.][for the correct identification of SNR4449-1]{mezcua13b}.
The ULX population is probably different in early- and late-type galaxies, due also to the different stellar population in the two classes of objects. In early-type galaxies, ULXs are consistent with being the luminous tail of the low-mass X-ray binary population \citep[e.g.][]{plotkin14}. 
The largest sample of ULXs comes from late-type galaxies \citep[see][and references therein]{swartz11}, correlating in general with Star Forming regions \citep[see e.g.][]{mineo12}.

The main questions that need to be answered can be summarised by:
$\bullet$ How can we determine the fundamental properties of ULXs? $\bullet$ Can we infer the Black Hole (BH) masses and mass-function? 
$\bullet$ And can we determine the growth of intermediate-mass BHs?

The nature of the ULX engine is still puzzling and continues to provide us with a strong theoretical
challenge, for instance how to grow massive BHs \citep{belczynski10}. 
In spiral galaxies, the number of ULXs has been found to correlate with the galaxy's global star formation rate, suggesting
that they are mostly high-mass X-ray binaries  \citep[HMXBs;][]{swartz04, swartz11, liu06}.  This is consistent with
studies that show good correlations of the star formation rate in a galaxy with its global hard X-ray luminosity \citep{ranalli03, persic07} and with the number of HMXBs \citep{grimm03, mineo12}, as confirmed by the study of the X-ray luminosity
function in ULX-rich galaxies \citep[e.g.][]{WT04, zezas06}.  
A significant fraction of ULXs may be fuelled by super-Eddington accretion onto stellar mass BHs \citep[5-20 M$_{\odot}$;][]{king01, king08}, with non-negligible beaming effects. On the other hand, a few X-ray point sources in distant galaxies have extreme
luminosities: L$_X > 10^{41}$ erg/s  \citep[the so-called HyperLuminous X-ray sources: HLX;][]{kaaret01, matsumoto01, wolter06, farrell09, sutton11}.  Such extreme luminosities \citep[and spectra, at least in the case of HLX-1 and M 82 X-1,][]{servillat11, FK10} can be easily explained by sub-Eddington accretion onto an intermediate-mass black hole (IMBH).  Black holes with stellar masses would need unreasonable beaming and/or super-Eddington accretion rates
to produce such high luminosities \citep[e.g.][]{swartz11}.  
It has also been proposed that bright persistent ULXs may contain the $\sim 20-80 M_{\odot}$ BH remnants of massive stars formed via direct collapse in a low metallicity environment \citep{mapelli09, ZR09}. A few observational
results seem to confirm this scenario \citep{pizzolato10, prestwich13, cseh14, ripamontiP}.  
The tentative identification of the orbital modulation of a ULX
optical counterpart \citep{liu09, zampieri12}, the discovery of a candidate
IMBH of over 500 M$_{\odot}$ in the galaxy ESO 243-49 \citep{farrell09, farrell12}, the detection of X-ray quasi-periodic oscillations (QPOs)  in a handful of ULX   \citep[e.g.][]{SM03, feng10, dewangan06, DS13}, and
the first tentative dynamical measure of the BH mass in M101 X-1 \citep{liu13} 
are only some examples of the fervid activity going on in the field.

The study of the timing properties of ULXs represents a promising way
to better understand their nature by comparison with the properties of
Galactic BH binaries (BHB). Galactic BHBs are among the brightest X-ray Galactic sources and have
been extensively studied over the last decades by several large multi-wavelength 
campaigns \citep[e.g.][]{RM06, fender09}. There are almost a
hundred known BHBs, both persistent and transient sources, and we have
acquired a fairly good knowledge of the physics underway in these
systems, where a soft accretion disk covered by a comptonizing cloud of hot electrons, 
responsible for the hard emission, explains well the observed emission. Different states observed in BHBs 
are probably associated with different accretion regimes. Certain state transitions
are associated with radio-emitting mass ejection events. 
This is supported by the observed correlation between aperiodic variability and X-ray flux
\citep{belloni10, munoz-darias11}.  This type of relation can
be used as a good tracer of the different accretion regimes in
Galactic sources.
In addition, a systematic analysis of spectral
variability of ULXs in the X-ray band with the help of the hardness-intensity and colour-colour diagrams
can provide a powerful tool to investigate the properties of their accretion flow also in comparison with the well-studied galactic BHBs. 
Studies in this direction were recently carried out by \citet{pintore14}
and \citet{sutton13} for a few bright, nearby ULXs with the best quality data. By organizing sources according to mean spectral 
colours and intensity, they have shown that the behaviour of the sources can be divided in two/three different groups with 
homogenous properties. Furthermore, they suggested that the observed spectral and temporal variabilities can be related 
to the mass accretion rate and inclination angle to these sources.
Understanding analogies and differences from Galactic BHBs is clearly important to 
shed light on the nature of ULXs, especially for very bright or hyper-luminous objects that may host IMBHs.

\section{ULXs in the radio band}

The radio band opens new possibilities not only by allowing a
multiband approach to the investigation of the ULX central energy source,
but also by determining the properties of the source
and its environment. We can distinguish between compact radio emission, extended emission from a jet, and shell-like emission from SNRs.
Radio spectral properties can help to distinguish between the various jet types
and to reveal the nature of the emitting source (eg. thermal or synchrotron emission).
Combined radio and X-ray data for a large sample of sources
can be compared with known classes of emitters, like AGNs, supernova
remnants, X-ray transients etc. Radio data also provide a means to
estimate source lifetimes and the energy associated with the equipartition assumption.

Until now, only a handful of ULXs have been detected in the radio, due
to the faintness of the sources relative to the sensitivity limits of the
current telescopes.  An observational VLA campaign aimed at detecting radio emission 
from a sample of 7 HLXs, that reached an rms of 0.01 mJy/beam yielded only 
one detection \citep{mezcua13c}.
Typically, observed ULXs have luminosities in the
radio band of the order L$_R = 10^{34-36}$ erg/s, from one to three orders
of magnitude lower than radio supernovae \citep{weiler02}.  The
radio to X-ray flux follows roughly a linear correlation, albeit with
a large scatter, similar to that derived for interacting supernovae
\citep{chevalier03}, but the difference in L$_R$ supports the hypothesis the
most ULXs are not supernovae.  Furthermore, ULX bubbles do not resemble
SNR: they carry at least two orders of magnitude more total energy, reach a size of hundreds of pc, 
have a different spectral shape, predominantly described by synchrotron emission
\citep{PM02, cseh12,  mezcua13c, mezcua14}.

The detected sources are mostly closer than 15 Mpc.  One notable exception is the radio counterpart of
HLX-1 at 95 Mpc, which has a flux density of 50 $\mu$Jy at cm wavelengths \citep{webb12} and was targeted just because it is a very luminous object \citep[but it may be the nucleus of a stripped dwarf or of a bulgeless satellite galaxy,][]{webb10, mapelli13}.  
A number of optical/UV nebulae have also been detected around nearby ULXs \citep{PM02, PM03}.  Three famous nebulae have been studied in the radio, Holmberg II X-1 \citep{miller04, cseh14}, NGC 5408 X-1 \citep{kaaret03} and IC 342 X-1 \citep{cseh12}: their radio spectra are consistent with optically thin synchrotron emission;
the radio nebulae can be compared with optical results and allow an estimate of the energetics of the systems.  All of these sources currently observed associated to a nebula are located within 5 Mpc \citep[see also][]{WZ13}. 
However, as the nebulae are resolved only for nearby ULXs, with the present instruments it is still
not known if the existence of the nebulae is ubiquitous in ULXs or if they
can be categorized in different classes. Preliminary results seem to imply that if they are
present they might be much fainter than the ones already observed \citep{mezcua13c}.  In some cases a compact core or 
resolved jets are observed \citep{ML11, mezcua13a, mezcua13c, mezcua14, cseh14}.
Indeed a thorough investigating of a complete sample of ULXs in this
respect is still lacking.
Furthermore, by assuming the emission of ULXs is due to accretion in a
sub-Eddington regime, similar to radiatively inefficient Galactic XRBs and AGNs, the radio
detection allows us to locate the ULXs in the fundamental plane (FP) of
sub-Eddington accreting black holes \citep{corbel03, gallo03, merloni03, falcke04} as defined by a
correlation between radio core (L$_R$) and X-ray (L$_X$) luminosity and
black hole mass ($M_{\rm BH}$), recently revised by  \citet{plotkin12} to 
%${\rm log\,} L_R = 0.39 {\rm log\,} L_X + 0.68 {\rm log\,} MBH + 16.61$.  Bonchi
${\rm log\,} L_{\rm X} = 1.45 \times {\rm log\,} L_{\rm R} - 0.88 \times {\rm log\,} M_{\rm BH} - 6.07$. 
This could be one of the
most viable option of determining or setting an upper limit to the mass of ULXs in the stated assumptions:
we can apply this when the sources are in the hard state and the radio emission comes from the compact jet.
For instance, by using the upper limits in the radio band and assuming a hard state in IC 342 X-1 with 
the FP  plane, \citet{cseh12} derived an upper limit on $M_{\mathrm{BH}}$ of 1000 $M_{\odot}$
\citep[see also][]{mezcua13c}. Further X-ray spectral studies confirmed this upper limit by setting more 
stringent mass constraints on the BH of 30-200 $M_{\odot}$ \citep{marlowe14}.
More recent findings nevertheless seem to exclude that the sources are found in the hard state, except
HLX-1, where a radiatively inefficient accretion flow (RIAF) was found \citep[NuSTAR observations by][]{bachetti13, walton13}.
%In IC342 X-1 the unresolved emission (or compact component) was interpreted as due to clump of emission from the bubble or due to flares (Cseh+ 2012, Marlowe+, 2014, submitted)
If a hard state is more commonly found for sources that are not persistent in the radio band,
then we should focus on repeated observations in order to find variable/transient sources
\citep[e.g. see review by][]{webb14}. 
%{\bf check 
Compact radio emission is observed in the hard state, while strong relativistic transient 
ejections are seen during state transitions, when the source goes from hard to soft  
state \citep[see][and references therein]{FB12, zhang13}.
%AR: I think this statement might be incorrect - compact jets are seen in hard states, and strong transient flares are seen in hard->soft transitions (at the X-ray maximum, normally dominated by a thermal component, the jet/radio is quenched).}

Planned and future surveys at centimetre-wavelenghts will repeatedly observe the sky, offering
thus the chance to study the variability of the sources and, possibly,
even to detect new sources based on their variability patterns. Such surveys include 
The Hunt for Dynamic and Explosive Radio
Transients conducted with Meerkat (a.k.a. Thunderkat
\footnote{http://www.ast.uct.ac.za/thunderkat/ThunderKAT.html}), and the
Australian Square Kilometre Array Pathfinder (ASKAP) Survey for
Variables and Slow Transients \citep{murphy13} that will reach a
continuum sensitivity of 47 $\mu$Jy/beam in 1 hr. The resolution
of the ASKAP survey (10$^{\prime\prime}$) will not allow the ULX to be resolved from the 
galaxy nucleus, while MeerKAT could in principle do this, if it reaches, as envisioned, a 50 mas resolution.
Also the JVLA is planning an all sky survey: VLASS\footnote{https://science.nrao.edu/science/surveys/vlass}.

The SKA, due to its large survey speed will allow monitoring of known radio
counterparts to determine both spectra and timing properties, and detection
of both new counterparts to known ULXs and possibly new ULXs. In addition, radio measurements will permit 
the estimate of the minimum energy associated with equipartition in the bubble 
(see \citet{kaaret03, miller04} for Ho II X-1 and \citet{mezcua13a} for N5457-X9).
%We could also investigate the presence of outflows. 
In case a jet can be identified to large distances and for faint sources,
we will be able to infer the mass of ULXs accreting at sub-Eddington rates.

\section{Measuring ULXs in the new era of advanced radio telescopes}
SKA observations of ULXs will study both the kinetic feedback of relativistic jets and the nebulous radio bubbles around their compact objects. Observations can be divided into measurements of either the steady extended structure or the possibly transient core emission. The former requires high-resolution observations to resolve the surrounding radio bubbles and to track the expansion of superluminal motion; the latter requires monitoring and rapid response measurements that will alert the community to new flaring events. As such, SKA-LOW is not a priority for studying ULXs, as the resolution is too low to resolve the bubbles and the transient decoherence time (period between recurrent events) may not be distinguishable due to the optical depth effect at long wavelengths. Therefore the discussion below will focus on the future GHz surveys from the SKA.
%{\it We do not consider the construction phase (eg. the 50\% outcome of SKA1) since the telescope will be not sensitive enough for our goals, while current instruments could be used, albeit with larger integration time [correct?] to the same effect.} {\bf ARP: I think we can do better here}
The study of ULXs will in general benefit from analysis of all the other emission sources in the galaxy, which is described in depth in \citet{beswick14}.

\subsection{Detecting `bubbles' surrounding ULXs}

A few example ULXs that have known associated radio bubbles are N5457-X9  $L_{\rm{1.6GHz}}=1\times10^{34}$~erg~s$^{-1}$ \citep{mezcua13a}, Holmberg II-X1 $L_{\rm{4.8GHz}} = 2.2\times10^{34}$~erg~s$^{-1}$ \citep{cseh14}, IC 342 X1 $L_{\rm{4.8 GHz}} = 1.8\times10^{35}$~erg~s$^{-1}$ \citep{cseh12}, plus the candidate micro-quasars in NGC7793 (S26) $L_{\rm{5.5GHz}}=2\times10^{35}$~erg~s$^{-1}$ \citep{soria10} and in M83 (MQ1) $L_{\rm{5.5GHz}}=2\times10^{35}$~erg~s$^{-1}$ \citep{soria14}. All have typical sizes of a few 10-100 pc across (corresponding to $\sim1-10^{\prime\prime}$ at a distance of a few Mpc). These bubbles have some similarities with the W50 nebula surrounding the Galactic microquasar SS433, observed to have a total flux of 70 Jy at 1.4 GHz, and assuming to have a distance of 5.5 kpc, would have a slightly lower luminosity of $3.5\times10^{33}$~erg~s$^{-1}$ compared to the ULX bubbles. 

\begin{figure}
  \centering
  \includegraphics[width=.75\textwidth]{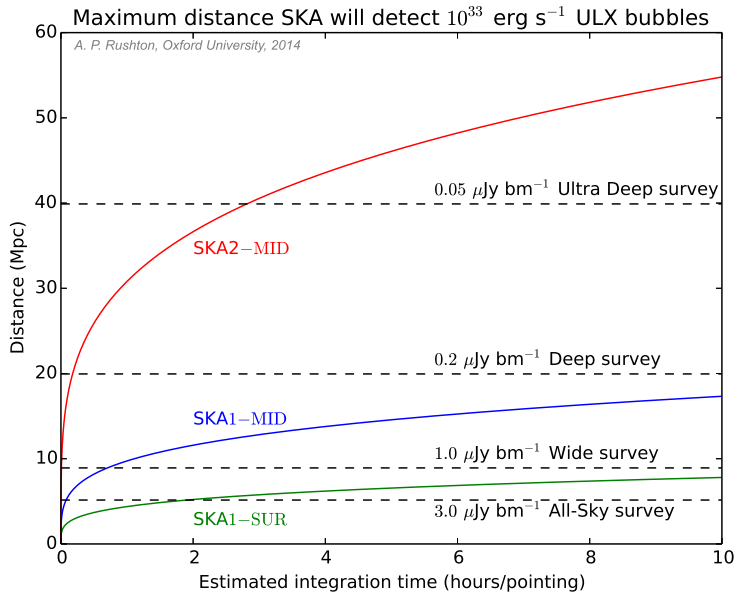}
  \caption{The maximum distance the SKA will detect ULX radio bubbles with a luminosity of $10^{33}$~erg~s$^{-1}$ as a function of integration time. The green, blue and red are the estimated integration time required to make a significant ($>7\sigma$) detection using SKA$1$-SUR, SKA$1$-MID and SKA$2$-MID respectively. The horizontal dashed lines show the maximum distance to which the bubbles can be significantly detected with four different surveys being planned with some of the reference continuum surveys discussed in \citet{PS15}. In particular we show the 1.4 GHz All-sky high resolution ($\leq 0.5^{\prime\prime}$) survey (3 $\mu$Jy bm$^{-1}$ rms), and the three tiers of the $0.5^{\prime\prime}$ resolution 1 GHz survey: `Wide' (1000-5000 deg$^2$, 1 $\mu$Jy bm$^{-1}$ rms), `Deep' (10-30 deg$^2$, 0.2 $\mu$Jy bm$^{-1}$ rms) and `Ultra Deep' (1 deg$^2$, 0.05 $\mu$Jy bm$^{-1}$ rms).}
  \label{fig:Survey}
\end{figure}

To measure all the ULX bubbles in the local universe, a reasonable luminosity limited survey should therefore reach a brightness of $L_{\rm{radio}} \leq 10^{33}$ erg~s$^{-1}$, in order to detect all the currently known types of diffused radio bubbles. This would correspond to a required flux detection of $S_{\rm{radio}} <835\times\nu_{GHz} D_{\rm{Mpc}}^{-2}~\mu$Jy bm$^{-1}$ or $S_{\rm{1.5GHz}} <550\times D_{\rm{Mpc}}^{-2}~\mu$Jy bm$^{-1}$. We also require a sub-arcsecond resolution to resolve the bubbles out to a distance of a few $\times10$ Mpc. In Figure~\ref{fig:Survey} we show the distance at which the SKA can make a $7 \sigma$ detection of this luminosity as a function of estimated integration time per field, for three different instances of SKA. The telescope parameters assume dual polarisation, with SEFD values of 391, 1630 and 16300 K m$^{-2}$ and instantaneous bandwidths of 500, 700 and 700 MHz for SKA$1$-SUR, SKA$1$-MID and SKA$2$-MID respectively. As shown in Figure~\ref{fig:Survey}, both SKA$1$-SUR and SKA$1$-MID quickly reach this detection limit for our Local Group of galaxies, while SKA$1$-SUR requires significantly more telescope time to detect ULX radio bubbles much further than $\sim5$ Mpc away. Moreover, SKA$1$-MID will significantly detect all radio bubbles within $<10$ Mpc by integrating for $\sim1$~hr per field (assuming rms noise of $\sim0.6~\mu$Jy bm$^{-1}$) and very deep SKA$1$-MID observations could detect ULXs bubbles out to $\sim18$~Mpc. Phase 2 of the SKA will double the distance of detecting ULX bubbles of a given luminosity with only modest telescope time per target. This will effectively increase the search volume by over an order of magnitude and very deep pointings could detect known radio bubbles out to $\sim55$~Mpc.

\subsection{Monitoring core emission from ULXs}

Whilst the steady radio emission from the surrounding ULX environment can be readily estimated from known sources, accreting sources (such as ULXs) may also produce core emission, directly related to the accretion process and possibly transient outbursts of varying luminosity and cadence. Outbursts from compact sources are associated with changes in the accreting conditions and a detailed coupling between the X-ray state (i.e. X-ray spectral hardness) and their radio producing jets has been well documented for X-ray binaries \citep{Corbel15}. Hitherto, there are only a few extragalactic transient sources that might originate from a super-Eddington stellar-mass or a sub-Eddington IMBH. The best example of transient behaviour comes from the Hyper-Luminous X-ray source HLX-1 in the galaxy ESO 243-49. \citet{webb12} showed that HLX-1 goes through regular outbursts to a flux density of 50--80~$\mu$Jy~bm$^{-1}$ (between 5--9 GHz) and assuming the host galaxy is at $\sim95$ Mpc, this gives a peak luminosity of $L_{\rm{5-9~GHz}}\sim2-8\times10^{36}$~erg~s$^{-1}$. HLX-1 goes through a semi-recurrent outburst with a cadence of around 1 year and an outburst duration of 100--200 days (although the peak emission only last a few days). Similarly, a radio transient in M82 reaches a flux density of $\sim1.5$ mJy~bm$^{-1}$  \citep{muxlow10} corresponding to a luminosity of $L_{\rm{5~GHz}}\sim1\times10^{35}$~erg s$^{-1}$ for a distance of 3.5 Mpc, although no clear X-ray emission is associated with this source, hence this object is not strictly a ULX (although it shares similar radio properties). The Ultra Luminous Infra-Red Galaxy Arp 220 (at a distance of $\sim77$ Mpc) also shows multiple recurrent radio transients with a flux density around few $100~\mu$Jy bm$^{-1}$ \citep{batejat12} corresponding to an apparent luminosity of $\sim7\times10^{36}$ erg~s$^{-1}$. Finally, the radio emission associated with a ULX in NCG 2276 produces a radio flux at 5 GHz of 1.43 mJy, which assuming a distance of 33.3 Mpc gives a radio luminosity of $\sim9.5\times10^{36}$~erg~s$^{-1}$ \citep{mezcua13c}; this emission could be either an extremely bright surrounding radio nebula or a very slowly evolving jet from an intermediate mass black hole, although we do not have yet a secure association with the galaxy \citep{wolter14}. Therefore, the brightest radio counterparts of ULXs \citep[or ULX-like sources like S26 in][]{soria10} appear to have a luminosity of up to $L_{\rm{radio}}\sim10^{37}$erg~s$^{-1}$, hence can already be detected with current instruments out to $\sim100$~Mpc. Identifying more of these `hyper-bright' off-nuclear radio sources may be a serendipitous by-product of the wide-field SKA surveys \citep[and Sect.~\ref{allsky} here]{Fender15}, however, high-resolution observations will be required to separate their locations from their galactic nuclei. Certainly, for target fields at >100 Mpc, SKA-VLBI will be needed as follow-up to resolve the few milli-arcsecond central core of the host galaxies \citep{Paragi15}.

\begin{figure} 
  \centering
  \includegraphics[width=.85\textwidth]{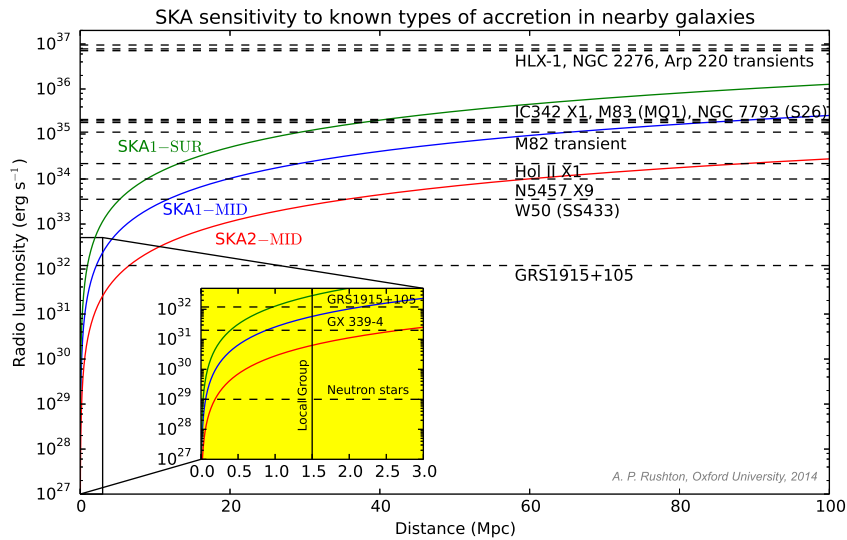}
  \caption{The sensitivity of the SKA to known transient sources in the nearby universe as function of distance. The green, blue and red are the estimated ($7 \sigma$) flux density detection limit of SKA$1$-SUR, SKA$1$-MID and SKA-MID respectively assuming a 1 hour integration per target. The horizontal dashed lines are the luminosities of known radio emitting ULXs and Galactic compact stellar sources.}
  \label{fig:Sensitivity}
\end{figure}

The high-sensitivity and fast survey speed of the SKA instruments will be particularly useful for finding slow transients and persistent core emission in the nearest galaxies. From observations of compact Galactic binaries (i.e. accreting neutron stars and stellar-mass black holes), we know that hard X-ray states produce a flat-spectrum radio jet. Transitions to softer X-ray states are associated with a steep spectrum radio flare (i.e. brighter at longer wavelengths), possibly due to the formation of a shock within the jet. We can therefore estimate the distance we will detect these outbursts if they are also associated with ULXs. For example, the brightest Galactic transient is Cygnus X-3, which can peak above 10 Jy bm$^{-1}$ \citep{miller-jones04}, and although the exact distance is uncertain (estimated between 7-10 kpc), this corresponds to around $L_{\rm{radio}}\approx5\times 10^{33}$ erg~s$^{-1}$ (peaking brighter than the entire W50/SS433 nebula!). GRS1915+105, the first superluminal source detected in the Galaxy, is also relatively bright with a peak flux of 250 mJy bm$^{-1}$ \citep{rushton10} and a distance of more than 10 kpc, which gives a $L_{\rm{radio}}\sim 10^{32}$ erg~s$^{-1}$. Finally, the radio luminosity directly emitted from the jet in SS433 produces a near persistent flux that flares up to $\sim 2.5$ Jy~bm$^{-1}$, although as the distance is only 5.5 kpc this is a corresponding luminosity of only $L_{\rm{1.0~GHz}}\approx 9\times 10^{31}$ erg~s$^{-1}$. As an example of the detection limit of the SKA, we show in Figure~\ref{fig:Sensitivity} the estimated $7\sigma$ detection limit of SKA to these known transients out to a distance of 100 Mpc for integrations of 1 hour.

%\section{Demonstrate the science outcomes that are enabled by the capabilities of a particular component of the SKA Phase 1 deployment}{\it APR: Is this now covered in section 3?}

%Aims: twofold. To survey uncharted regions of the sky, To reach further in detecting ULXs. (Distance TBD). To observe at selected intervals known and bright objects to determine their behaviour as a class. 

%Pb: confusion limit to be determined: how far out can we look for ULXs with SKA; e.g. <what is the expected sensitivity 
%and angular resolution. 
%LR range of detected sources. 

In case of radiatively inefficient accretion flow, which usually means accretion at low or very low Eddington rates, the Fundamental Plane method can be applied. Although it is questionable to use the FP as a general tool, in those cases in which it can be applied it yields an independent estimate of the BH mass. For instance, a 100 M$_{\odot}$ at low Eddington rate would produce $L_{\rm{X}}=1\times10^{40}$~erg~s$^{-1}$ and consequently $L_{\rm{R}} \sim 1 \times10^{33}$~erg~s$^{-1}$ according to \citet{plotkin12}.

\subsection{How to exploit the All-Sky Surveys for ULXs}
\label{allsky}

The SKA will be used to make all-sky surveys and it is therefore interesting to quantify the ULX discovery potential of such surveys.
Following \citet{PS15} we consider two 1.4 GHz all-sky ($3\pi$; the total sky visible to the SKA) radio-continuum reference surveys for SKA in its phase 1 (SKA$1$): one characterised by $2\,\mu$Jy~bm$^{-1}$ rms, $2^{\prime\prime}$ resolution; the
other one by 3 $\mu$Jy bm$^{-1}$ rms and $0.5^{\prime\prime}$ resolution. With such rms
sensitivities we are able to detect the faintest ($>10^{33}$ erg s$^{-1}$) ULX only in the very local Universe
(only up to a distance of 5-6 Mpc), while the brightest ULX ($>10^{35-36}$
erg s$^{-1}$) can be detected up to 50-150 Mpc. As far as angular
resolution is concerned, $2^{\prime\prime}$ is more than enough to be able to
distinguish ULX candidates from other point sources at the galaxy center, in the majority of situations,
as it allows us to resolve scales of the order of 50-500 pc at 5-50 Mpc.

Assuming a factor 10 increase in sensitivity, we can anticipate that SKA
all-sky surveys will be able to push ULX blind searches up to distances of
150-200 Mpc for source luminosities of $10^{35}$ erg s$^{-1}$.  Sub-arcsec angular resolutions will be needed to resolve
physical scales of $\sim 500$ pc.

The first and easiest task would be to identify radio counterparts or upper limits for the known ULXs in current and future lists. We emphasise that the ULXs with a detected radio counterpart are at this time only a small fraction of the ULX population. We envisage that after having measured a large enough number of radio data for ULXs we will be able to better characterise the sources properties and define a parameter space that will allow the detection of new objects directly from the radio observations. New accreting sources could be identified for instance from the recurrent ejection of jets, while the determination of their bolometric luminosity would come a-posteriori from multi-wavelength follow-up observations. Current X-ray facilities as XMM-Newton and Chandra are expected to last many years and are providing catalogs of X-ray sources to compare with the future radio sky. Athena has been selected by ESA in the framework of the Cosmic Vision 2015-2025 plan, and is currently planned to a launch in the late 2020s. Other instruments have been proposed to have either a sharper resolution (like Smart-X) or timing properties (like LOFT) and will be at work in synergy with the SKA if approved.
% Athena: http://www.the-athena-x-ray-observatory.eu/
% Smart-X; http://smart-x.cfa.harvard.edu/
% LOFT: http://www.isdc.unige.ch/loft/index.php/the-loft-mission

\subsection{Searching for ULXs during the SKA early science phase}

The estimated sensitivity of surveys presented in this chapter assume the completion of phase 1 and 2 of the SKA, in order to detect the maximum number of ULXs and produce the largest statistical sample of these compact objects in the radio band. Owing to the time-critical nature of these accreting sources, the early science phase of the SKA will be crucially important and will provide a longer monitoring period for the brightest sources. The cadence for decoherence between outbursts is typically months to years (and sometimes a few decades) and regular monitoring of the nearest galaxies should commence immediately. 

Known and bright sources such as HLX-1 could be significantly detected within only a few minutes integration time with 50\% of SKA1-MID and about 30 mins. using 50\% of SKA1-SUR; therefore, 50\% SKA1-MID could be used to regularly monitor the known transient ULXs and be triggered when outbursts are detected at other wavelengths, whilst 50\% SKA1-SUR could blindly search for serendipitous events. Also, during the early science phase, the SKA will be the most sensitive radio telescope in the Southern hemisphere and can be used to make targeted searches for radio bubbles from the nearest ULXs.

\acknowledgments
It is a pleasure to acknowledge useful discussions with Robert Beswick and Sara Motta. F.P., A.W. and L.Z. acknowledge financial support from the INAF research grant PRIN-2011-1 (\lq\lq Challenging Ultraluminous X-ray sources: chasing their black holes and formation pathways'') and financial contribution from the agreement ASI-INAF I/037/12/0.

\end{document}